\documentclass{PoS}

\title{Extracting scattering parameters using the isospin chemical potential}

\ShortTitle{Extracting scattering parameters using the isospin chemical potential}

\author{\speaker{Michael I. Buchoff}%
        \thanks{In collaboration with Paulo Bedaque and Brian Tiburzi.}
        \thanks{This work was partially supported by the U.S. Dept. of Energy under grant number DEFG02-93ER-40762.}
         \thanks{Preprint numbers:  UMCP-DOE-40762-490, LLNL-PROC-461612}\\
       $^1$Maryland Center for Fundamental Physics, Department of Physics\\
University of Maryland, College Park, MD 20742 \\
       $^2$ Physical Sciences Directorate, Lawrence Livermore National Laboratory \\Livermore, California 94550, USA\\
       E-mail: \email{buchoff1@llnl.gov}}

\abstract{Hadronic scattering mediated through the strong interaction has been an 
area of great interest for both theory and experiment. Recently, lattice 
QCD calculations of scattering processes from first principles have seen 
remarkable progress, including multiple calculations of two-pion 
scattering lengths that agree with experiment to within a few percent. 
However, there exists a certain class of scattering processes, such as 
pion-nucleon scattering, that contain annihilation diagrams, which are often
prohibitively expensive to simulate on the lattice. In this talk, I will 
present a method to extract certain parameters from this class of 
scattering processes by employing an isospin chemical potential, which 
can be simulated on the lattice as a result of its positive-definite 
fermion determinant.
}

\FullConference{The XXVIII International Symposium on Lattice Field Theory\\
                June 14-19,2010\\
                Villasimius, Sardinia Italy}

\begin{document}
\def\a{{\alpha}}
\def\b{{\beta}}
\def\d{{\delta}}
\def\D{{\Delta}}
\def\X{{\Xi}}
\def\e{{\varepsilon}}
\def\g{{\gamma}}
\def\G{{\Gamma}}
\def\k{{\kappa}}
\def\l{{\lambda}}
\def\L{{\Lambda}}
\def\m{{\mu}}
\def\n{{\nu}}
\def\o{{\omega}}
\def\O{{\Omega}}
\def\S{{\Sigma}}
\def\s{{\sigma}}
\def\th{{\theta}}

\def\ol#1{{\overline{#1}}}

\def\Aslash{A\hskip-0.45em /}
\def\Dslash{D\hskip-0.65em /}
\def\Dtslash{\tilde{D} \hskip-0.65em /}

\def\CPT{{$\chi$PT}}
\def\QCPT{{Q$\chi$PT}}
\def\PQCPT{{PQ$\chi$PT}}
\def\tr{\text{tr}}
\def\str{\text{str}}
\def\diag{\text{diag}}
\def\order{{\mathcal O}}

\def\cF{{\mathcal F}}
\def\cS{{\mathcal S}}
\def\cC{{\mathcal C}}
\def\cB{{\mathcal B}}
\def\cT{{\mathcal T}}
\def\cQ{{\mathcal Q}}
\def\cL{{\mathcal L}}
\def\cO{{\mathcal O}}
\def\cA{{\mathcal A}}
\def\cQ{{\mathcal Q}}
\def\cR{{\mathcal R}}
\def\cH{{\mathcal H}}
\def\cW{{\mathcal W}}
\def\cM{{\mathcal M}}
\def\cD{{\mathcal D}}
\def\cN{{\mathcal N}}
\def\cP{{\mathcal P}}
\def\cK{{\mathcal K}}
\def\Qt{{\tilde{Q}}}
\def\Dt{{\tilde{D}}}
\def\St{{\tilde{\Sigma}}}
\def\cBt{{\tilde{\mathcal{B}}}}
\def\cDt{{\tilde{\mathcal{D}}}}
\def\cTt{{\tilde{\mathcal{T}}}}
\def\cMt{{\tilde{\mathcal{M}}}}
\def\At{{\tilde{A}}}
\def\cNt{{\tilde{\mathcal{N}}}}
\def\cOt{{\tilde{\mathcal{O}}}}
\def\cPt{{\tilde{\mathcal{P}}}}
\def\cI{{\mathcal{I}}}
\def\cJ{{\mathcal{J}}}

\def\eqref#1{{(\ref{#1})}}

\section{Introduction}


Within the past decade, there has been substantial progress in understanding scattering phenomena from first-priciple lattice QCD calculations through analyzing two-hadron correlation functions.  Several meson-meson calculations, most notably the I=2 $\pi\pi$ scattering channel, agree with experimental calculations to within a few percent \cite{Yamazaki:2004qb,Beane:2005rj,Beane:2007xs,Feng:2009ij}.  In addition to the meson-meson channel, the multi-baryon channel has received a great deal of attention \cite{Beane:2009py} and continues to show considerable progress. 

One set of processes that has not received as much attention are baryon-meson processes.  These processes are needed not only to understand long-range interactions of nucleons through pion exchange, but kaon-nucleon interactions play a pivotal role in understanding potential kaon condensation in neutron stars \cite{Kaplan:1986yq}.  While calculations have been preformed for several meson-nucleon scattering processes \cite{Torok:2009dg}, there exist a class of processes that have eluded these calculations; namely, calculations with annihilation diagrams.   While immense progress has been made in recent years due to advancements in algorithms and the use of GPUs,  calculation of annihilation diagrams are often prohibitively expensive requiring all-to-all propagators.  

In this work, we present an alternative method of extracting these parameters through the use of an isospin chemical potential.  This method, proposed in Ref. \cite{Bedaque:2009yh}, takes advantage of the dynamically generated pion condensate when the isospin chemical potential reaches a certain critical value.  Then, through the aid of heavy baryon effective field theory in its region of applicability, the relevant scattering parameters can be extracted by performing baryon spectroscopy in the presence of this pion condensate. 

\section{Annihilation Contributions in Meson-Baryon Processes}
\begin{figure}[b] 
   \centering
   \begin{tabular}{cccc}
  \includegraphics[width=3in]{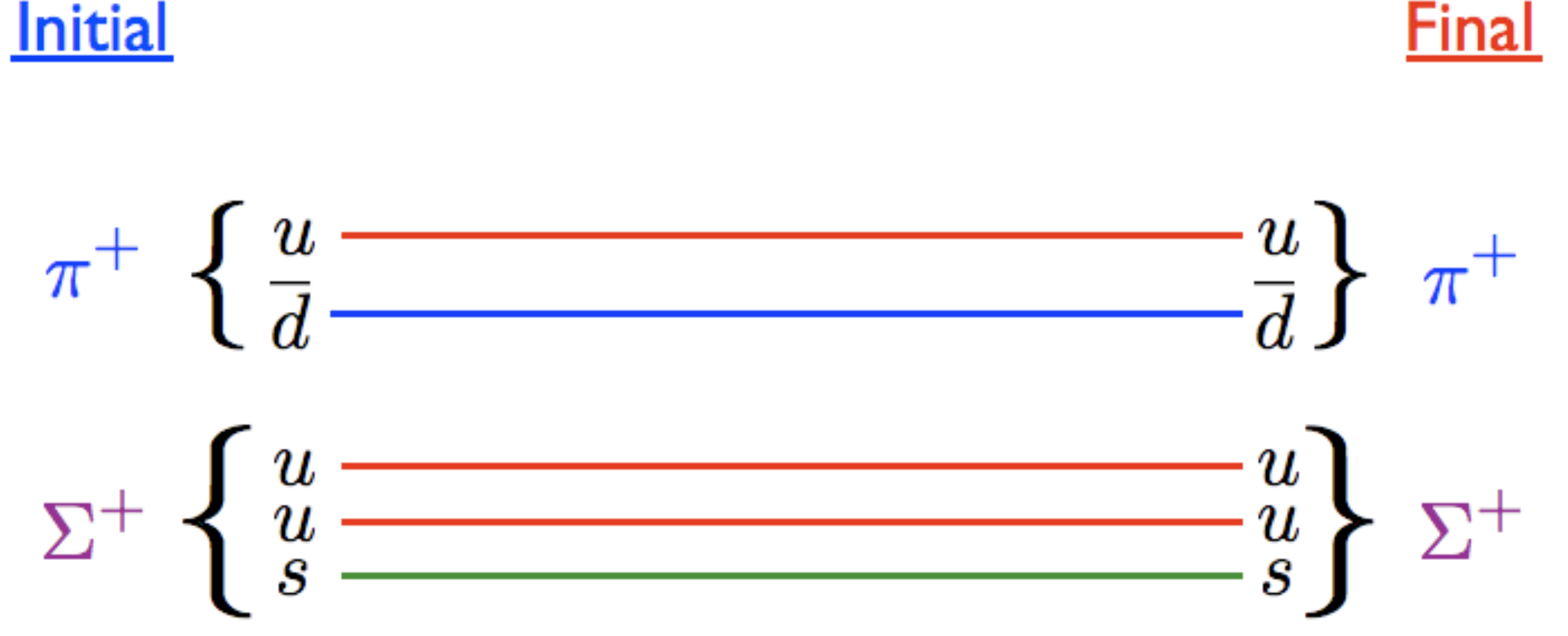}  &  & & \includegraphics[width=3in]{Sig_Pi_1_better} 
  \end{tabular}
   \caption{Examples of the ``non-annihilation" contributions to the $\pi^+\Sigma^+$ process.}
   \label{fig:Sig_Pi}
\end{figure}

As mentioned previously, currently lattice methods of extracting scattering information rely on two (or more) hadron correlation functions.  For meson-baryon processes, this consists of putting a meson and baryon operator in the source and sink, performing the relevant contractions, and then summing over all possible final states.  For some meson-baryon processes, such as $\pi^+ \Sigma^+$, there are only ``non-annihilation"  processes, as shown in Fig.~\ref{fig:Sig_Pi}.  As a result, only a point-to-all propagation (effectively one column of the matrix) is necessary to extract the relevant information, and these methods were preformed in Ref.~\cite{Torok:2009dg}.

\begin{figure}[t] 
   \centering
   \includegraphics[width=4in]{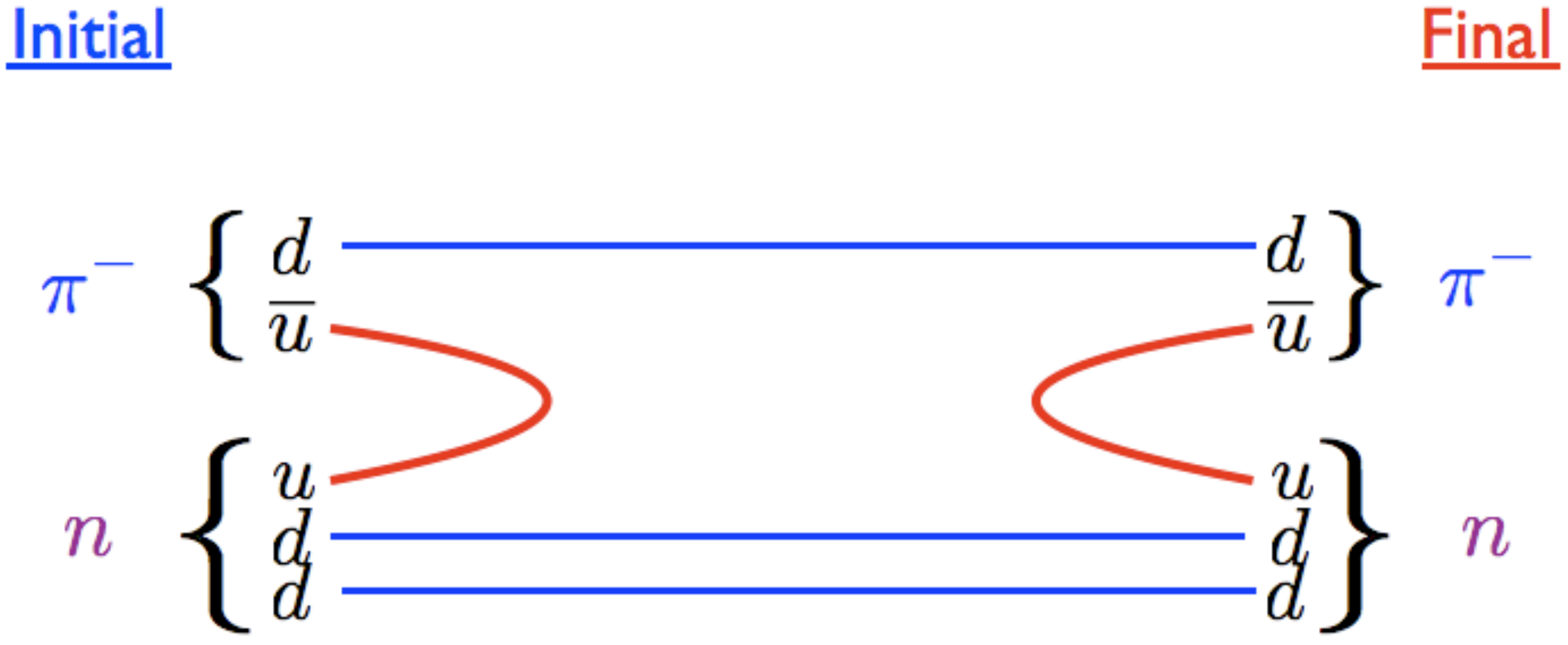}
   \caption{Examples of the ``annihilation" contributions to the $\pi^+ n$ process.}
   \label{fig:Pi_n}
\end{figure}

However, there exist another class of scattering processes, such as $\pi N$ or $K^- N$, that contain annihilation processes, which can have a significant effect on the scattering process (the $\Lambda$-resonance in $K^- N$ scattering is an example of this).  These effects, as illustrated in Fig.~\ref{fig:Pi_n}, require an all-to-all propagation (whole matrix is required), which is often very expensive to calculate using brute force methods.  In light of this issue, we explore a different calculation, that only requires baryon spectroscopy in the chiral regime.

\section{Taking advantage of condensates}
\begin{figure}[t] 
   \centering
   \includegraphics[width=5in]{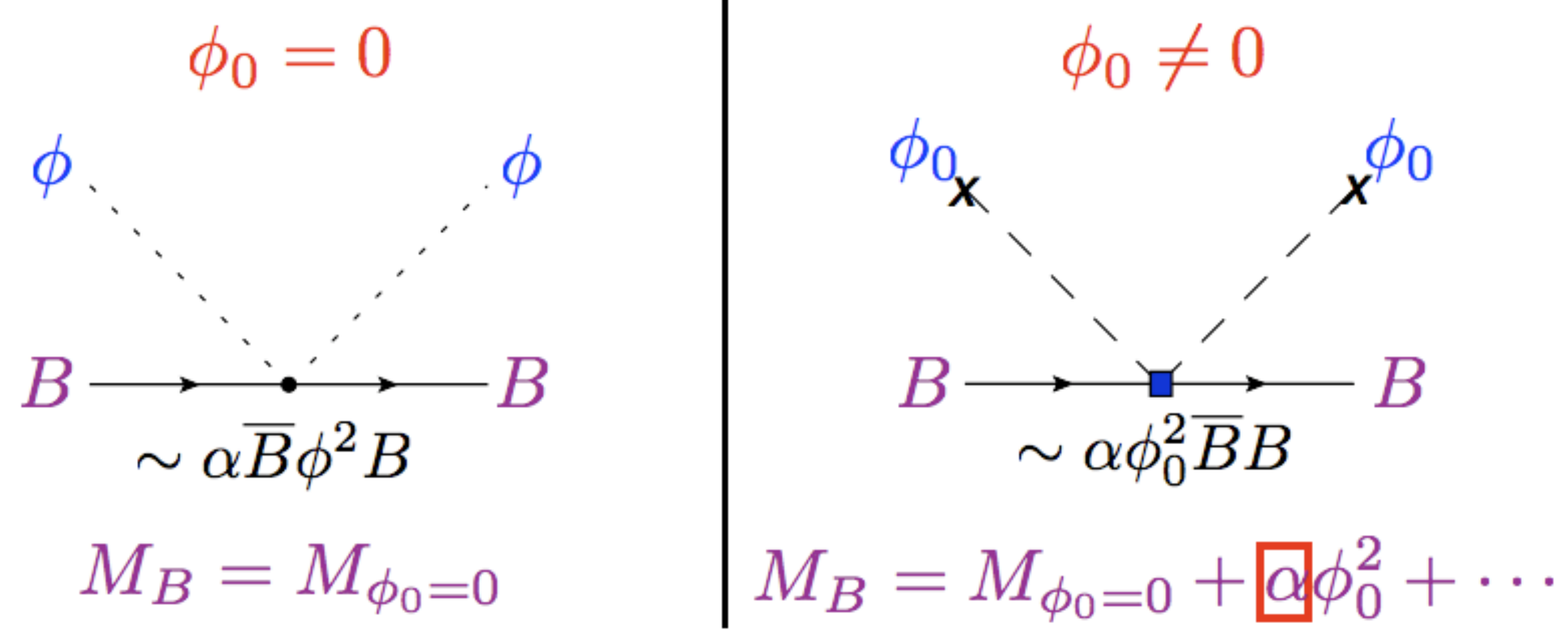}
   \caption{Diagrammatic illustration of how the condensate can shift the mass. By taking the difference of the masses calculated in both pictures and varying $\phi_0$, one can extract the scattering parameter of interest.}
   \label{fig:condense_mass}
\end{figure}

It has been well established for many years that condensates can be used to shift or generate masses.  The Higgs mechanism in particle physics is an example of this process as the Higgs vacuum expectation value gives masses to the vector bosons.  So could this process help extract scattering processes on the lattice?  For simplicity, we explore a system where a scalar without flavor indices interacts with a baryon.  The term in the Lagrangian for this process is given by
\begin{equation}
\label{ eq:L_BB_phi}
\mathcal{L} = \alpha \overline{B} \phi^2 B + \cdots.
\end{equation}
In most low-energy processes that we are interested in, the value of $\alpha$ is a function of the non-perterbative physics that can only be extracted through lattice calculations or experiment.  As mentioned previously, this process may contain annihilation diagrams that may render the lattice calculations prohibitively expensive.  However, if we were to dynamically form a scalar condensate, namely  $\phi(x) = \phi_0 + \tilde{\phi}(x)$, the term in the Lagrangian becomes
\begin{equation}
\label{ eq:L_BB_phi}
\mathcal{L} = \alpha\phi_0^2  \overline{B} B + \cdots.
\end{equation}
Since $\alpha\phi_0^2$ is a constant, this term can be thought of as a shift of the baryon mass, as shown in Fig.~\ref{fig:condense_mass}.  To be more precise, if the baryon mass without the condensate is $M_{\phi_0 = 0}$ then the baryon mass in the presence of the condensate would be $M_{\phi_0 = 0} + \alpha\phi_0^2 + \cdots$.  If we knew and could control the value of $\phi_0$, we could, in principle, extract the parameter $\alpha$ simply by taking differences of the baryon masses.

So the two questions that remain are, first, do we have a mechanism to dynamically generate a condensate, and, two, if we generate this condensate, do we have a method to systematically account for its effects?  Indeed, there are methods that allow both of these questions to be answered favorably.  The isospin chemical potential can generate a pion condensate and effective field theory gives us a controlled tool to see how the mass will shift as a function of this condensate.

\section{Isospin chemical potential}
One well known problem that occurs when studying systems at finite baryon density is the fermion sign problem.  When the baryon chemical potential is non-zero, the Euclidean action no longer has a positive definite determinant, which prohibitively hinders the usual Monte Carlo methods for lattice simulations.   As a result, there has been much effort studying finite density systems without a fermion sign problem.  One such example is the isospin chemical potential.

The action for the isospin chemical potential for the two flavor theory in Minkowski space is given by
\begin{equation} \label{eq:L_Iso}
\cL 
=
\ol \psi 
\left[ 
i 
\left(
\Dslash 
+
i
\mu
\gamma_0
\frac{\tau^3}{2}
\right)
-
m_q
\right]
\psi
,\end{equation}
By inspection (after the Euclidean rotation), one can see that the usual condition of $\gamma_5$-Hermiticity is not satisfied.  However, a different condition $\tau_1\gamma_5$-hermaticity\footnote{$\tau_2 \gamma_5$-Hermiticity is equally valid for this Lagrangian.}  is satisfied, which ensures a positive definite determinant:
\begin{equation}
\label{eq:pos_def}
\tau_1\gamma_5 M \gamma_5 \tau_1 = M^\dag \quad \rightarrow \quad \det M = \det M^\dag.
\end{equation}  
This feature has enabled multiple lattice calculations at finite isospin density \cite{Kogut:2002zg,Kogut:2004zg,deForcrand:2007uz}.  It is also worth pointing out the symmetries of this action.  When $m = \mu=0$, the action processes the usual $SU(2)_L \otimes SU(2)_R$ flavor symmetry which spontaneously breaks to $SU(2)_V$.  When $m \neq 0$ and $\mu \neq 0$, the action processes a $U(1)_L \otimes U(1)_R$ flavor symmetry which breaks, both explicitly and spontaneously 
to $U(1)_V$.

In addition to not having a prohibitive fermion sign problem, the isospin chemical potential leads to a pion condensation phase \cite{Kogut:2002zg,Kogut:2004zg,deForcrand:2007uz,Son:2000xc,Son:2000by}.  This condensation phase, which breaks the $U(1)_V$ symmetry spontaneously, occurs when the chemical potential reaches a certain critical value.  Basic arguments imply that this transition would occur when $\mu = m_\pi$, since there are no lighter $I=1$ particles that could be generated for smaller chemical potential values.  However, for a theory where the quark mass is light enough for chiral perturbation theory to be valid, this can be shown explicitly.

\section{Connections to effective field theory}
The way you map the isospin chemical potential term to the chiral effective field theory is to treat the chemical potential term as a gauge field, namely
\begin{equation}\label{iso_gauge}
\mathbb{V}_\mu = \mu \frac{\tau^3}{2}\delta_{\mu,0},
\end{equation}
while obeying the power counting $p^2 \sim m_\pi^2 \sim \mu^2$.  The resulting chiral Lagrangian is given by 
\begin{equation} \label{eq:CPT}
\cL
=
\frac{f^2}{8}\Big[
\mathbf{tr} ( D_\mu U D^\mu U^\dagger
)
+
2 \lambda 
\mathbf{tr}(M^\dagger U + U^\dagger M
)
\Big]
,\end{equation}
where $D_\mu U= \partial_\mu U +i[\mathbb{V}_\mu,U]$ and $U$ is an $SU(2)$ matrix of the pseudo-Goldstone bosons and can be written as 
\begin{equation}
\label{eq:U }
U = e^{i \alpha \mathbf{\hat{\pi}}\cdot \mathbf{\tau}} = \cos \alpha + i \mathbf{\hat{\pi}}\cdot \mathbf{\tau} \sin \alpha.
\end{equation}
Minimizing the potential yields the result \cite{Son:2000xc,Son:2000by}
\begin{equation}
\label{eq:chem_val }
\cos \alpha = \frac{m_\pi^2}{\mu^2}.
\end{equation}
There are several concepts that can be taken away from this equation.  First, when $\mu < m_\pi$, the system dynamically behaves the same way as the $\mu = 0$ scenario and when $\mu > m_\pi$, a pion condensate is formed and the chiral condensate gets altered.   Second, both the chiral condensate and pion condensate can be given in terms of the parameter $\alpha$.  If we add a small isospin breaking term to ensure that the spontaneous breaking of the pion condensate is in the $\tau_2$ direction (a process that is conveyed in more detail in Ref.~\cite{Bedaque:2009yh}), the values of the condensates are given by
\begin{eqnarray}
\langle \ol \psi \psi \rangle 
&=&
f^2 \lambda \cos \a,
\\
i \langle \ol \psi \tau^2 \gamma_5 \psi \rangle
&=&
f^2 \lambda \sin \a
.\end{eqnarray}
Throughout the rest of this note, the results will be given in terms of $\cos \alpha$ and $\sin \alpha$.  However, thanks to these relations, we can relate $\cos \alpha$ to the value of the chiral condensate and $\sin \alpha$ to the value of the pion condensate.

To reiterate the goal of this work, we want to find a relation between the mass shift of a hadron in the presence of an isospin chemical potential to the value of the condensate, chemical potential, and scattering parameters.  Thus, if we know the value of the condensate and chemical potential, along with the mass shift calculated on the lattice, we should be able to extract several elusive scattering parameters.  It should also be reiterated that the measurements of these mass shifts do not require all-to-all propagation and can be calculated using the usual spectroscopy methods.  In order to find this analytic relation between baryon mass shifts and the condensate/chemical potential, heavy baryon chiral perturbation theory is employed \cite{Bernard:1992qa}.     It should be emphasized at this point that for this analytic relation to be valid, the effective field theory must not break down.  In other words, these relations will only be useful for pion masses and chemical potential values in the chiral regime.  

We will now present the results and implications for the two flavor nucleon calculation\footnote{For the complete analysis, along with the calculations including the strange quark, see Ref.~\cite{Bedaque:2009yh}, using the two flavor hyperon analysis in Ref.~\cite{Tiburzi:2008bk} and performing the relevant matching.}. The nucleon mass dependence on the chemical potential and condensates is given by (in terms of the coefficients from the heavy baryon lagrangian in Ref.~\cite{Bernard:1992qa,Frink:2004ic})
\begin{equation}\label{M_N}
M_N 
=
M_N^{(0)} 
- 
\mu_I \cos \a \frac{\tau^3}{2}
+
4 c_1 \left( m_\pi^2 \cos \a + \lambda \epsilon \sin \a \right)
+
\left( c_2 - \frac{g_A^2}{8 M}  + c_3 \right) \mu_I^2 \sin^2 \a,
\end{equation}
where the low energy constant $c_1$ is the ``sigma term" and $c_2$ and $c_3$ determine the scattering length for pion-nucleon scattering.  The $\epsilon$ in this equation represents the small isospin breaking term needed to point the pion condensate in a particular direction in flavor space.  As mentioned previously, this is explained in Ref.~\cite{Bedaque:2009yh}.  The key message to take away from this equation is that the chemical potential (and consequently, the condensates) acts as an additional ``knob" for extracting low-energy constants.  Starting with the sigma term, $c_1$, which is still difficult to determine precisely on the lattice \cite{WalkerLoud:2008bp}, by varying $\mu$, the value of $\cos \alpha$ will vary, and $c_1$ can be extracted at a fixed pion mass.  This is a particularly interesting result as you can extract information about chiral physics if the pion mass were at the physical point.  The scattering parameters $c_2$ and $c_3$ can also be extracted by plotting the mass shift of the nucleon vs the  chemical potential times the pion condensate.  It should be pointed out, that this method alone cannot disentangle $c_2$ and $c_3$.  However, the implementation of twisted boundary conditions could separate these terms.  
Additionally, it is worth mentioning the approximate size of the mass shift and whether or not it can be reasonably extracted.  If we use the estimate of $c_1 \sim .9 \; GeV^{-1}$ \cite{Frink:2004ic} it is not difficult to see the nucleon mass will need to be determined from .5\% to 5\% depending on the pion mass (smaller pion masses require better resolution in the spectroscopy calculation).

\section{Conclusions and Future Applications}
The primary motivation for exploring the effect of isospin chemical potentials on hardronic mass shifts was to extract information about scattering processes that require calculation of annihilation diagrams on the lattice.  This all-to-all process is very expensive and this particular method allows one to move the issues from the propagator calculations to generating multiple configurations at different chemical potentials (in a similar way lattice calculations are generated at multiple mass values).  However, this approach can be more appropriately though of as an additional ``knob" on the theory that one can extract chiral information from.  In Ref.~\cite{Bedaque:2009yh}, the baryon-meson channel was the primary focus, but a similar calculation could be preformed for the meson-meson channel where calculating meson masses in the presence of  an isospin chemical potential will lead to meson-meson scattering.

This begs the question of whether or not other potentially interesting information can be extracted from lattices with an isospin chemical potential.    Since pion condensation is one of the possible phases that could exist within neutron stars, it would be worthwhile to understand the low-energy phenomena in this state.  Some potentially interesting calculations include heavy quark potentials, two meson phase shifts, etc.  Also, it would be enlightening to explore the thermodynamics of these systems.  If one can vary the temperature and isospin chemical potential, it should be possible to find a region of small fugacity when the virial expansion is valid.  The coefficients in these expansions contain information about three and four meson interactions and could be directly extracted from the lattice using these methods.


\begin{thebibliography}{99}
\bibitem{Yamazaki:2004qb}
T.~Yamazaki {\it et al.}  [CP-PACS Collaboration],
``I = 2 Pi Pi Scattering Phase Shift with Two Flavors of $O(A)$ Improved   Dynamical Quarks,''
Phys.\ Rev.\  D {\bf 70} (2004) 074513
[arXiv:hep-lat/0402025].

\bibitem{Beane:2005rj}
S.~R.~Beane, P.~F.~Bedaque, K.~Orginos and M.~J.~Savage  [NPLQCD
Collaboration],
``I = 2 Pi Pi Scattering from Fully-Dynamical Mixed-Action Lattice QCD,''
Phys.\ Rev.\  D {\bf 73} (2006) 054503
[arXiv:hep-lat/0506013].


\bibitem{Beane:2007xs}
S.~R.~Beane, T.~C.~Luu, K.~Orginos, A.~Parreno, M.~J.~Savage, A.~Torok and A.~Walker-Loud,
``Precise Determination of the I=2 Pipi Scattering Length from Mixed-Action   Lattice QCD,''
Phys.\ Rev.\  D {\bf 77} (2008) 014505
[arXiv:0706.3026 [hep-lat]].

\bibitem{Feng:2009ij}
X.~Feng, K.~Jansen and D.~B.~Renner,
``The Pi+ Pi+ Scattering Length from Maximally Twisted Mass Lattice QCD,''
Phys.\ Lett.\  B {\bf 684} (2010) 268
[arXiv:0909.3255 [hep-lat]].

\bibitem{Beane:2009py}
S.~R.~Beane {\it et al.}  [NPLQCD Collaboration],
``High Statistics Analysis Using Anisotropic Clover Lattices: (Iii)   Baryon-Baryon Interactions,''
Phys.\ Rev.\  D {\bf 81} (2010) 054505
[arXiv:0912.4243 [hep-lat]].

\bibitem{Kaplan:1986yq}
D.~B.~Kaplan and A.~E.~Nelson,
``Strange Goings on in Dense Nucleonic Matter,''
Phys.\ Lett.\  B {\bf 175} (1986) 57.


\bibitem{Torok:2009dg}
A.~Torok {\it et al.},
``Meson-Baryon Scattering Lengths from Mixed-Action Lattice QCD,''
Phys.\ Rev.\  D {\bf 81} (2010) 074506
[arXiv:0907.1913 [hep-lat]].


\bibitem{Bedaque:2009yh}
P.~F.~Bedaque, M.~I.~Buchoff and B.~C.~Tiburzi,
``Meson-Baryon Scattering Parameters from Lattice QCD with an Isospin   Chemical Potential,''
Phys.\ Rev.\  D {\bf 80} (2009) 114501
[arXiv:0910.4595 [hep-lat]].


\bibitem{Kogut:2002zg}
J.~B.~Kogut and D.~K.~Sinclair,
``Lattice QCD at Finite Isospin Density at Zero and Finite Temperature,''
Phys.\ Rev.\  D {\bf 66} (2002) 034505
[arXiv:hep-lat/0202028].


\bibitem{Kogut:2004zg}
J.~B.~Kogut and D.~K.~Sinclair,
``The Finite Temperature Transition for 2-Flavor Lattice QCD at Finite   Isospin Density,''
Phys.\ Rev.\  D {\bf 70} (2004) 094501
[arXiv:hep-lat/0407027].

\bibitem{deForcrand:2007uz}
P.~de Forcrand, M.~A.~Stephanov and U.~Wenger,
``On the Phase Diagram of QCD at Finite Isospin Density,''
PoS {\bf LAT2007} (2007) 237
[arXiv:0711.0023 [hep-lat]].


\bibitem{Son:2000xc}
D.~T.~Son and M.~A.~Stephanov,
``QCD at Finite Isospin Density,''
Phys.\ Rev.\ Lett.\  {\bf 86} (2001) 592
[arXiv:hep-ph/0005225].


\bibitem{Son:2000by}
D.~T.~Son and M.~A.~Stephanov,
``QCD at Finite Isospin Density: from Pion to Quark Antiquark   Condensation,''
Phys.\ Atom.\ Nucl.\  {\bf 64} (2001) 834
[Yad.\ Fiz.\  {\bf 64} (2001) 899]
[arXiv:hep-ph/0011365].


\bibitem{Bernard:1992qa}
V.~Bernard, N.~Kaiser, J.~Kambor and U.~G.~Meissner,
``Chiral Structure of the Nucleon,''
Nucl.\ Phys.\  B {\bf 388} (1992) 315.

\bibitem{Frink:2004ic}
  M.~Frink and U.~G.~Meissner,
  JHEP {\bf 0407}, 028 (2004)
  [arXiv:hep-lat/0404018].

\bibitem{Tiburzi:2008bk}
  B.~C.~Tiburzi and A.~Walker-Loud,
  Phys.\ Lett.\  B {\bf 669}, 246 (2008)
  [arXiv:0808.0482 [nucl-th]].


\bibitem{WalkerLoud:2008bp}
  A.~Walker-Loud {\it et al.},
  Phys.\ Rev.\  D {\bf 79}, 054502 (2009)
  [arXiv:0806.4549 [hep-lat]].


\end{thebibliography}
\end{document}